\documentclass[review]{elsarticle}

\usepackage{lineno,hyperref}
\modulolinenumbers[5]

\usepackage{microtype}
\usepackage{amsmath,amsthm,amssymb}
\usepackage{graphicx,epstopdf,epsfig}

\usepackage{multirow}
\usepackage[caption=false,font=footnotesize]{subfig}
\usepackage[linesnumbered,ruled]{algorithm2e}
\usepackage{xcolor}
\usepackage{array}
\usepackage{float}
\usepackage{enumerate}
\usepackage{diagbox}
\usepackage{tabularx,booktabs}

\usepackage{booktabs,multirow}
\def\S{\mathbf{S}}

\journal{ Journal of Visual Communication and Image Representation}

\begin{document}

\begin{frontmatter}

\title{Image Downscaling via Co-occurrence Learning}

\author{Sanjay Ghosh}
\address{Department of Radiology and Biomedical Imaging, \\
	University of California San Francisco, USA.}

\author{Arpan Garai}
\address{Department of Computer Science \& Engineering, \\
	Indian Institute of Technology Delhi, India.}

%
%

\begin{abstract}
Image downscaling is one of the widely used operations in image processing and computer graphics. It was recently
demonstrated in the literature that kernel-based convolutional filters could be modified to develop efficient image downscaling algorithms. In this work, we present a new downscaling technique which is based on kernel-based image filtering concept. We propose to use pairwise co-occurrence similarity of the pixelpairs as the range kernel similarity in the filtering operation. The co-occurrence of the pixel-pair is learned directly from the input image. This co-occurrence learning is performed in a neighborhood based fashion all over the image. The proposed method
can preserve the high-frequency structures, which were present in the input image, into the downscaled image. The idea is further extended to the case of fractions factor of downscaling. The resulting images retain visually-important details and do not suffer from  edge-blurring artifact. We demonstrate the effectiveness of our proposed approach with extensive experiments on a large number of images downscaled with various downscaling factors.
\end{abstract}


\end{frontmatter}


\section{Introduction}
\label{sec:intro}
The evolving display technology has brought a substantive increase in display devices offering a wide range of image resolutions.
These days, we rarely view a photo at its original resolution. It happens quite frequently that we reduce its original size (megapixel) to much smaller dimensions to be viewed on.
To convert an images to a different resolution, it is desirable to develop efficient image downscaling and upsampling (super-resolution) techniques. 
\textcolor{blue}{Downscaling is nothing but a fundamental  compression process by which the size of the high-resolution (HR) input image is reduced to generate a low-resolution (LR) output image \cite{occorsio2022lagrange}.}
The goal of image downscaling is to produce a lower resolution image by removing pixels while preserving the original image content as much as possible \cite{duchon1979lanczos}.
In addition, downscaling is also an integral step in image coding (downscaling is performed before the compression step).
In similar direction, downscaling of frames are performed in SD/HD video coding. \textcolor{blue}{Image downscaling is also intensively used for fast browsing  or sharing purposes \cite{chen2020learned}}.
There is a substantial need to develop efficient algorithms for image downscaling.

Most of the existing downscaling algorithms perform some form of linear filtering on the image. In particular, the filtering is performed using convolution (with a kernel) before subsampling and subsequent reconstruction, following the sampling theorem \cite{shannon1984communication, duchon1979lanczos}, {\cite{mitchell1988reconstruction}}. The key objective in these works was to minimize the aliasing artifacts. However, the downscaled images obtained using these strategies are usually over-smoothened. Sometimes the perceptually important details and features cannot be retained. This is primarily because of the non-adaptive nature of the filtering kernel, which is independent of the image content. 
The same phenomenon is observed in more recent image interpolation techniques \cite{thevenaz2000interpolation, nehab2011generalized,trentacoste2011blur, samadani2008representative}.
Motivated by the working principle of bilateral filtering \cite{Tomasi1998}, the authors in \cite{kopf2013content} proposed to adapt the kernel shapes to local image patches. We note that bilateral filtering is an example of kernel filtering \cite{milanfar2013tour,huang2016fast}. Recently, another kernel based method was introduced in \cite{weber2016rapid}, where a bilateral-like similarity measure was used in the kernel filtering process. We note that kernel filtering is routinely used in various image quality enhancement tasks \cite{garcia2015unified,ren2008bilateral} including denoising \cite{gabiger2013fpga,dabhade2017reconfigurable}, sharpening \cite{ghosh2019saliency}, inpainting \cite{wong2008nonlocal} etc. In this work, we explore it's capacity for image downscaling.

\noindent
\textbf{Contributions:}
In this paper, we present a new algorithm for image downscaling. The working of the proposed approach is built on the idea of kernel filtering. 

\begin{itemize}
	\item We introduce a new measure for the range similarity in the filtering step. In particular, we learn pixel-pair similarity within a local neighborhood from the (whole) input image. This pair-wise co-occurrence proximity is then used in computing the final downscaled image via some form of kernel filtering. \textcolor{blue}{For the sake of completeness, we note that the overall computation complexity of the proposed method is same as a classical kernel filtering using pixel statistics such as bilateral filtering.}
	
	\item We further extend our method for performing downscaling with non-integer scaling factor. This is an attractive aspect of makes our method. 
  In real-world scenarios, the scale factor should be arbitrary, such that original images with various resolutions need to meet the resolution restriction. We note that many of the existing methods cannot be used for fractional scaling.
\end{itemize}

\noindent
\textbf{Organization}. 
The rest of the paper is organized as follows. 
We present a review of the literature on image downscaling in Section \ref{sec:related}.
The proposed downscaling algorithm is described in Section \ref{sec:algo}, where we also provide a complexity analysis of it. In Section \ref{sec:results}, we report several experiments, exhibiting very competitive results. To demonstrate the effectiveness of our proposal, we performed exhaustive downscaling experiments with a wide range of image classes and different downscaling factor.  Finally, we conclude in Section \ref{sec:conc}.

\section{Related Work}
\label{sec:related}

The classical techniques for image downscaling were originated from the idea of sampling theory \cite{shannon1984communication}. The main streamline of these techniques is to - first filter the image to blur the edges
and then subsample the intermediate image (as per the desirable factor) to obtain the final downscaled image. The purpose of applying this lowpass filter is to prevent aliasing in the downscaled image. Finding a suitable lowpass filter along with its kernel parameters remains to be a prominent question for further research in this direction. An alternative
approach for image downscaling  without filtering is to construct the downscaled image by directly optimizing some form of proximity (similarity measure) between the original and downscaled images.
We now present a detailed review of the literature on image downscaling algorithms. In doing so, we group them into two classes depending on whether they are - filtering based or optimization based. In addition, few deep learning based approaches {\cite{saeedan2018detail, hou2017deep, kim2018task, sun2020learned}} have been introduced recently.

\subsection{Filtering based methods}
The key step in most of the traditional algorithms is kernel filtering which aims to suppress the high frequency components in order to avoid aliasing artifacts in the downscaled image \cite{duchon1979lanczos,mitchell1988reconstruction}. The elementary filters are - box, bicubic, and Lanczos filter \cite{duchon1979lanczos}. These simple kernel filtering methods fail to preserve edges; thus the downscaled image suffers from over-smoothing artifacts \cite{zhou2017scale}.
To retain some of the important details in the downscaled images, authors in \cite{triggs2001empirical, park2020edge} proposed a sophisticated kernel filter design.  
To reduce ringing and over-smoothing artifacts, authors in \cite{nehab2011generalized} applied interpolation on image downscaling by adding an extra construction step.
A joint bilateral filtering based approach was introduced in \cite{kopf2013content}. 
For each output pixel, a suitable filter kernel was adaptively estimated by considering both spatial and color variances
of the local region around that pixel.
Thus, in contrast to pure segmentation, each input pixel exhibits a weighted contribution to each of the output pixels.
Additional care was taken to avoid excessive deformation of the input image and smooth edges.
Another joint bilateral filtering based method was proposed in \cite{weber2016rapid} where the
range kernel of the bilateral filter favors differences in local pixel neighborhoods. The similarity between two pixels are evaluated in such a way that the high-frequency details are preserved in the downscaled image. 
The core idea is based on information theoretic intuition that a piece of data deviating from its neighbors usually contains more valuable information. 
Moreover, it is believed that the human visual system \cite{beghdadi2013survey} is more sensitive to such piece of information in an image. This  fast downscaling algorithm \cite{weber2016rapid} was particularly suited for large images and videos.
{\color{blue} In \cite{occorsio2022lagrange}, the continuous scaling to resize an image is performed using  Lagrange polynomial interpolation. 
Authors in \cite{occorsio2022image} applied bivariate polynomial sampling based interpolation for resizing.
}

\subsection{Optimization methods}
There are approaches which model image downscaling as an optimization problem. The downscaling problem was formulated to  maximize the structural similarity index (SSIM)  between the input and corresponding downscaled image \cite{oztireli2015perceptually}. 
The solution leads to a non-linear filter, which computes local luminance and contrast measures on the input image and a smoothed version of it.
The global optimization framework could be able to render most of fine details. However, the spatial distribution of pixels within each patch is usually not considered in the SSIM index computation. Therefore, this  technique \cite{oztireli2015perceptually} sometimes fails to handle structured patterns, thus resulting in aliasing artifacts.
An interpolation-dependent downscaling method was proposed in \cite{zhang2011interpolation} where the objective of the optimization was set to minimize the sum of square errors between the input image and the one interpolated from the corresponding downsampled image.
The problem of downscaling was posed as a $L_0$ regularized optimization in \cite{liu2018L0}. It consists of two $L_0$ priors. One of the priors was intended to enforce the fact that there is an inverse square proportionality between the number of edge pixels of the downscaled image and the downscaling factor. Another prior was used to minimize the sparsity of the downsampling matrix.
Image downscaling was performed by remapping high frequencies to the representable range of the downsampled spectrum  in \cite{gastal2017spectral}. The input image was first decomposed into a set of spatially-localized non-harmonic waves; and then the frequency remapping was modeled as an optimization problem. {\color{blue} In \cite{cho2021learning}, a phase is estimated from the high and low resolution image and a patch class is estimated based on edge and direction analysis in a region. The image is downscaled using  the kernel which is  trained by optimising these parameters.}

{\color{blue}
\subsection{Deep learning based methods}
In recent times, a number of deep learning based method are already proposed for image downscaling. In \cite{saeedan2018detail}, an adaptive pooling method is proposed. In this method, the weights used in the averaging process are kept in such a way that the pixel that have higher intensity difference in the input image are passes through the next step. Hou et al. proposed a deep convolutional neural network (CNN) based down-scaling approach in \cite{hou2017deep}. In \cite{hou2017deep}, the images are down-scaled to $\frac{1}{4}$ of the size of the input image (which is considered as $256\times 256$). The CNN introduced in  \cite{hou2017deep} generates a number of intermediate image and finally selects the image that have similar spatial correlations with the input image as output image. In \cite{kim2018task}, the bicubic downsampled version of the input image is used as guide image, whereas, in \cite{li2018learning}, the bicubic downsampled version of the input image is added directly to produce a better result. A resampler network have been introduced in \cite{sun2020learned} for image downsampling. Authors in \cite{xing2022scale} introduced a universal encoder-decoder styled deep network. Note that the downscaling task is dealt by the encoder 
network  while the decoder network is for the upscaling task. Thus, the encoder-decoder is trained jointly to address their duality. Unlike most deep neural network based approaches which perform integer resizing, the method in \cite{chen2021convolutional} performs non-integer downsampling using deep-NN. It is performed by a CNN based down scaling and followed by a bicubic upscaling.   
}

\subsection{Relationship Between the Proposed Method and Existing Schemes}
Our proposed method belong to the class of kernel based methods. The core idea was motivated by the spirit of joint bilateral filtering based approaches in \cite{kopf2013content, weber2016rapid}. Unlike classical bilateral smoothing, the goal of an efficient image downscaling method is to successfully retain the fine details. Therefore, from a methodological point of view, the challenge is to come up with an appropriate/optimized kernel filtering scheme which would retain the contrast of the image as best possible. With this direction of thought, authors in \cite{ weber2016rapid} introduced a form of inverse-bilateral kernel. Authors in \cite{kopf2013content} performed a special form of bilateral combination of two Gaussian kernels defined over space and color. In contrary, we propose a novel kernel filtering scheme based on co-occurrence statistics of pixel pairs in the input image.

\section{Proposed Approach}
\label{sec:algo}

The proposed convolutional filter based method determines the filter weights such that important details (pixel-pair) are retained in the processed/output image. Similar to \cite{weber2016rapid}, we first construct an image of the desired dimension  by performing box filtering, followed by Gaussian convolution. We refer this intermediate image as guide image.

Suppose the dimension of the input image $f$ is $M \times N$. For simplicity, let us assume that we are interested in $d \times d$ folded downscaling. {\color{blue} Here, d is the downscaling factor. } Thus the dimension of the downscaled image $g$ is $m \times n$, where $m = M/d$ and $n = N/d$. In general, a $d\times d$ folded downscaling means that each pixel in $g$  is generated from a patch of size $d \times d$ in the input image $f$. 

In the box filtering step, we perform an arithmetic averaging of all the pixels in each patch and assign that value  as the intensity of downscaled intermediate image (of size $m \times n$). This filtering operation is the step where we obtain a crude estimate of the downscaled image.

In the subsequent steps, we refine the pixel values using kernel filtering. Let the intermediate image after box filtering be referred as $f_d$.
The {\color{blue} $i^{th}$ pixel of the} output image, $g$ of size $m \times n$  is given by:

\begin{equation}
	g(i)  = \frac{\sum_{\j\in \Omega(i) }  f(j)  C \big(f_{d}(i) , f(j) \big) }{\sum_{\j\in \Omega(i) } C \big(f_{d}(i) , f(j) \big)},
	\label{eq:main}
\end{equation}
where $C \big(f_{d}(i) , f(j) \big)$ is the co-occurrence frequency {\color{blue} (which is explained in section \ref{sec_co_occ})} of the intensity pair $\big(f_{d}(i) , f(j) \big)$ in the input image $f$. Note that $f_{d}(i) $ is the intensity level of the intermediate guide image (which could have non-integer values as well). For using the co-occurrence frequency we round-off $f_{d}(i) $ to nearest integer. The two-dimensional neighborhood in (\ref{eq:main}) is $\Omega (i)  = i + \Omega_d$, where $\Omega_d = [-d, d]^2$ {\color{blue} which indicates the 2D neighborhood of the ROI of the input image. }We note that this neighborhood $\Omega_d$ is with respect to the input image $f$. Therefore the filtering operation in (\ref{eq:main}) is not similar to those traditional kernel filtering on images. Instead, this can be considered as a special type of guided filtering where the pixel intensities in $f_d$ is updated using one particular neighborhood similarity of $f$. {\color{blue} For RGB images, we perform the similar operation for each channel.  In other words, each band is downscaled separately by using different co-occurrence profile to the respective channel.} 
We refer our proposed method as image downscaling via co-occurrence learning (IDCL).

\begin{figure*}[htp!]
\centering
\subfloat[$k = 2$.]{\includegraphics[width=0.49\linewidth]{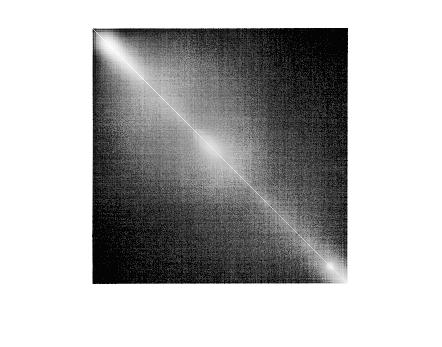}}
\hfill
\subfloat[$k = 3$.]{\includegraphics[width=0.49\linewidth]{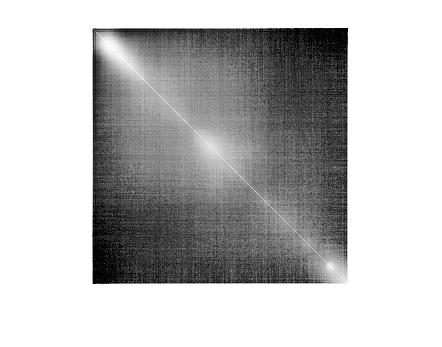}}
\caption{Co-occurrence map for two different kernel width parameter $k$ values. The downscaling factor $d = 2$ in both cases.}
\label{fig:detail}
\end{figure*}

\subsection{Nonlocal Co-occurrence}\label{sec_co_occ}
\begin{figure}
	\centering
	\subfloat[Input.]{\includegraphics[width=0.5\linewidth]{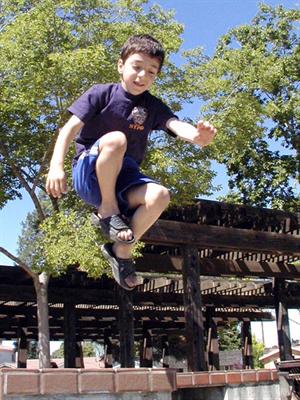}} \\
	\subfloat[k = 2.]{\includegraphics[width=0.25\linewidth]{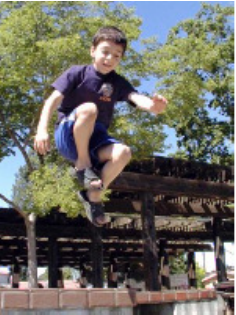}} 
	\hspace{2mm}
	\subfloat[k = 3.]{\includegraphics[width=0.25\linewidth]{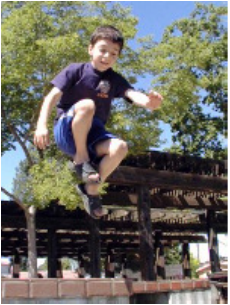}} 
	\hspace{2mm}
	\subfloat[k = 5.]{\includegraphics[width=0.25\linewidth]{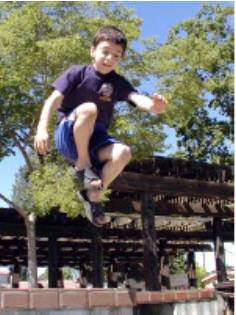}} 
	\caption{Effects of the model parameter $k$ for downscaling factor $2$. The input image of $(300 \times 400)$ is downscaled by scale of 2. All output images are visually indistinguishable. {The PSNR/SSIM of the pairs \{(b, c), (c, d), \& (d, b)\} are \{48.73/99.97, 43.75/98.23, \& 48.24/99.89\} respectively.}}
	\label{fig:img02}
\end{figure}

The pair-wise co-occurrence in (\ref{eq:main}) is computed solely from the input image. 
We define co-occurrence similarity as following:
\begin{equation}
	C(a, b) = \sum_{\i} \sum_{j \in \S(i) }[f(i)  = a] [f(j)  = b],
	\label{eq:co-occurence}
\end{equation}
where [$\cdot$] equals $1$ if the expression inside the brackets is true and $0$ otherwise. 
The neighborhood in  (\ref{eq:co-occurence}) is $\S(i)  = i + \mathbb{S}$, where $\mathbb{S} = [-k, k]^2$. Similar to the rationale in \cite{weber2016rapid}, we set, $k \geq d$. For simplicity, we assume that the intensities are rounded off to nearest integer. For a 8-bit image, the possible values of $a$ and $b$ are $\{0, 255\}$. 
We note that $C(\cdot, \cdot)$ stores all co-occurrence pixel intensity-pair across the image over the neighborhood $\mathbb{S}$. In other words, $C(\cdot, \cdot)$ captures the co-occurrence profile of the original image which is parameterized by $k$.
We refer $h$ as the nonlocal co-occurrence matrix of the input image. The term ``nonlocal'' is to indicate the fact that we compute the cumulative co-occurrence of each possible pair $(a, b)$ throughout the whole image over restricted neighborhood $\mathbb{S}$ of size $(k \times k)$ around each pixel in the input image. {\color{blue} Also, in other words, the non-local co-occurrence values indicate the measurement of the similarity between the intensities of the pixels in the neighbourhood $(k \times k)$.} We refer this step of our proposal as \emph{co-occurrence learning}. 
\textcolor{blue}{We show an example of co-occurrence maps for the setting in Figure \ref{fig:detail}. The co-occurrence maps for $ k = 2$ and $ k = 3$ learned from the input image in Figure \ref{fig:img02}(a). We mention that Matlab in-built function \textit{tonemap} is used for the display purpose.}

\begin{algorithm}[t!]
	\DontPrintSemicolon
	\KwIn{Input image $f$ of size $(M \times N)$; \\ \hspace{1cm} 
		Neighborhood parameter $k$.}
	\KwOut{Co-occurrence profile $C$ of size $(256 \times 256)$.} 
	Set $C$ as a zero-matrix of size $(256 \times 256)$\;
	\For{$i = 1, \ldots, (M \times N)$} {
		$a = \lfloor f_d(i)  \rceil$  \quad \% $\lfloor \cdot \rceil$ stands for round-of operation\;
		\For{$j \in  \S(i) $}{
			$b = \lfloor f(j)  \rceil $ \; 
			$C(a, b) = C(a, b) + 1$\;
		}
	}
	\caption{Co-occurrence learning.}
	\label{algo:co-occur}
\end{algorithm}
\begin{algorithm}[t!]
	\DontPrintSemicolon
	\KwIn{Input image $f$ of size $(M \times N)$; \quad
		intermediate image $f_d$ of size $(m \times n)$; \quad co-occurrence profile $C$ of size $(256 \times 256)$.}
	\KwOut{Downscaled image $g$ of size $m \times n$.}
	\For{$i = 1, \ldots, (m \times n)$} {
		Set $P = 0$ and $Q = 0$\;
		$a = \lfloor f_d(i)  \rceil$   \quad \% $\lfloor \cdot \rceil$ stands for round-of operation\;
		\For{$j \in  \Omega(i) $}{
			$b = \lfloor f(j)  \rceil $\; 
			$w = C( a, b)$ \quad \% obtained from \textbf{Algorithm \ref{algo:co-occur}} \;
			$P = P + w f(j) $\;
			$Q = Q + w $\;
		}
		$g(i)  = P/Q$\;
	}
	\caption{Kernel filtering.}
	\label{algo:filtering}
\end{algorithm}

\begin{algorithm}[t!]
	\DontPrintSemicolon
	\KwIn{Input image $f$ of size $(M \times N)$;  
		intermediate image $f_d$ of size $(m \times n)$, where $m = ceil(M/d)$ and $n = ceil(N/d)$; \quad co-occurrence profile $C$ of size $(256 \times 256)$.}
	\KwOut{Downscaled image $g$ of size $m \times n$.}
	\For{$i = 1, \ldots, (m \times n)$} {
		Set $P = 0$ and $Q = 0$\;
		$a = \lfloor f_d(i)  \rceil$   \quad \% $\lfloor \cdot \rceil$ stands for round-of operation\;
		\For{$j \in  \Omega(i) $}{
            $\tilde{d} = ceil(d)$ \quad \% $ceil$ stands for ceiling operator\;
            $\Omega(i) = [(i -  \ \tilde{d}), (i + \Tilde{d})]$ \quad \% 2D neighborhood around $i$.\;
			$b = \lfloor f(j)  \rceil $\; 
			$w =  C( a, b)$ \quad \% obtained from \textbf{Algorithm \ref{algo:co-occur}} \;
			$P = P + w f(j) $\;
			$Q = Q + w $\;
		}
		$g(i)  = P/Q$\;
	}
	\caption{Kernel filtering with fractional downscaling factor $d$.}
	\label{algo:filtering_frac}
\end{algorithm}

\subsection{Computation Complexity}
The proposed method IDCL consists of three steps:
\begin{enumerate}[(i)]
	\item Box filtering.
	\item Learning the co-occurrence profile (\textbf{Algorithm \ref{algo:co-occur}}).
	\item Kernel filtering (\textbf{Algorithm \ref{algo:filtering}}).
\end{enumerate}
For completeness, we provide a computation complexity analysis in terms of input pixels ($M \times N$), downscaling factor $d$, and model parameter $k$. 
It was shown in \cite{crow1984summed} that box filtering of an image can be performed at constant time complexity (independent of the kernel width). So, the filtering of an image of ($M \times N$) pixels would require $O(M N)$ computations. 

In the co-occurrence learning step in Algorithm \ref{algo:co-occur}, we need to perform $O(k^2)$ comparisons for each pixel in the input image. So the total complexity is $O(MNk^2)$. Finally, there are $O(d^2)$ computations for each pixel in the intermediate image as reported in Algorithm \ref{algo:filtering}. The intermediate image contains total  $(M N)/d^2$ pixels. 
The overall computations required for downscaling an image of  ($M \times N$) pixels by a factor of $d$  is: 
%
\begin{equation}
	\begin{split}
		& = O(MN) + O(MN k^2) + \frac{MN}{d^2} O(d^2) \\
		& = O(MN) + O(MN k^2) + O(MN) \\
		& = O(MN k^2).
	\end{split}
\end{equation}
Alternatively, the per-pixel complexity with reference to input image is $O(k^2)$. In most experiments, we used $k = d$; therefore the per-pixel computation is $O(d^2)$. We note that \cite{weber2016rapid} has very similar complexity as our method except of the co-occurrence learning step. Given $d$ (downscaling factor in each spatial dimesion), our method would require only a little more computation time than \cite{weber2016rapid}. On a different note, we mention that the idea of co-occurrence fairly holds for gray-scale images. However, for downscaling a color image, we need to process each of the three color channels separately.

\subsection{Properties of Co-occurrence Kernel}
The co-occurrence kernel $h$ exhibits the following properties:
\begin{enumerate}
	\item Positivity: $h(a, b) \geq 0, \ \forall (a,b)$.
	\item Symmetric: $h(a, b) = h(b,a)$.
	\item Positive definite: For any feasible collection of points $t_i$, $i = 1, \ldots, n$, the Gram matrix with elements $h_{i,j} = h(t_i, t_j)$ is positive definite matrix.
\end{enumerate}
Further, it can also be shown that this co-occurrence kernel satisfies the Cauchy-Schwartz inequality $h^2(a,b) \leq h(a, a) h(b,b)$.

\subsection{Non-integer Downscaling Factor}
We often require fractional ratios in practical real-world applications. 
\textcolor{blue}{In the current era of social media, we happen to resize/downscale an image while viewing on our own display device. Also in case of mobile photography, we happen to see all the captured images with varying resolutions on one display of fixed resolution.}
It is interesting to note that our method could be extended for fraction downscaling ratios. 

Our proposed method consists of two steps: \textcolor{blue}{ Co-occurence learning (Algorithm \ref{algo:co-occur} with $k = ceil(d)$) and kernel filtering (Algorithm \ref{algo:filtering_frac})}. For a given downscaling ratio, we need to decide kernel parameter $k$ in step 1. We choose to set $k = \text{ceil} (d)$, where $ceil$ stands for ceiling operator. In the filtering step, we choose a neighborhood/patch on the input image of size $k \times k$ for each pixel in the downscaled image. 

\section{Experimental Results}
\label{sec:results}

\subsection{Parameters}
In our algorithm, there is only one parameter to control, $k$. It basically determines the size of the square neighborhood $\S$ while building the similarity matrix in \eqref{eq:co-occurence}. Note that each of the possible pair occurs in the aggregation step in \eqref{eq:main} should also be considered during the construction of the co-occurrence matrix in  \eqref{eq:co-occurence}. 
This basic observation leads to $k \geq d$. In a practical scenario of downscaling, the downscaling factor $d$ is a user-defined quantity. Whereas, $k$ is the algorithmic parameter to be set with reference to $d$. In Figure \ref{fig:img02}, we present a visual result to demonstrate the effect of various possible $k$ for a fixed $d$. In this experiment, we used $d = 2$ and $k = \{2, 3, 5\}$. We notice that all three downscaled images are visually indistinguishable. In other words, the proposed method IDCL is less sensitive to $k$.

\begin{table}[!h]
	\centering
	\caption{Comparioson of computation time.}
	\label{table}
	\addtolength{\tabcolsep}{7pt}
	\begin{tabular}{||c | c | c ||}
		\hline 
		{Lanczos} \cite{duchon1979lanczos}  & {DPID} \cite{weber2016rapid}  & {IDCL}   \\ \hline 
		0.006 sec.  & 0.55 sec.     &0.86 sec.   \\ \hline
	\end{tabular}
	\label{tab:time}
\end{table}

\noindent
\textbf{Runtime Comparison}: To demonstrate the effectiveness of our method, we present a comparison on computation time with \cite{duchon1979lanczos} and \cite{weber2016rapid}. In Table \ref{tab:time}, we show the results for downscaling a $(256 \times 256)$ color image by factor ($k = 2$) on \texttt{Matlab 2018a} in \texttt{Intel(R)} \texttt{3.50GHz} machine. We note that the simplest method in a \cite{duchon1979lanczos} takes the least time. Our method IDCL takes slightly more time than \cite{weber2016rapid}. This additional time is spent in learning the co-occurrence map as stated in Algorithm \ref{algo:co-occur}.

\begin{figure}
	\centering
	\subfloat[Input.]{\includegraphics[width=0.56\linewidth]{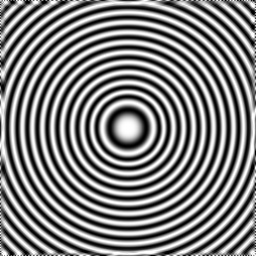}} \\
	\subfloat[DPID.]{\includegraphics[width=0.3\linewidth]{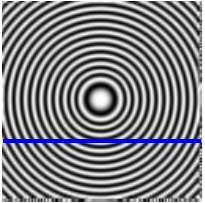}}
	\hspace{3mm} 
	\subfloat[IDCL.]{\includegraphics[width=0.3\linewidth]{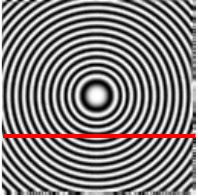}} \\
	\subfloat[The intensity values of the highlighted pixels.]{\includegraphics[width=0.98\linewidth]{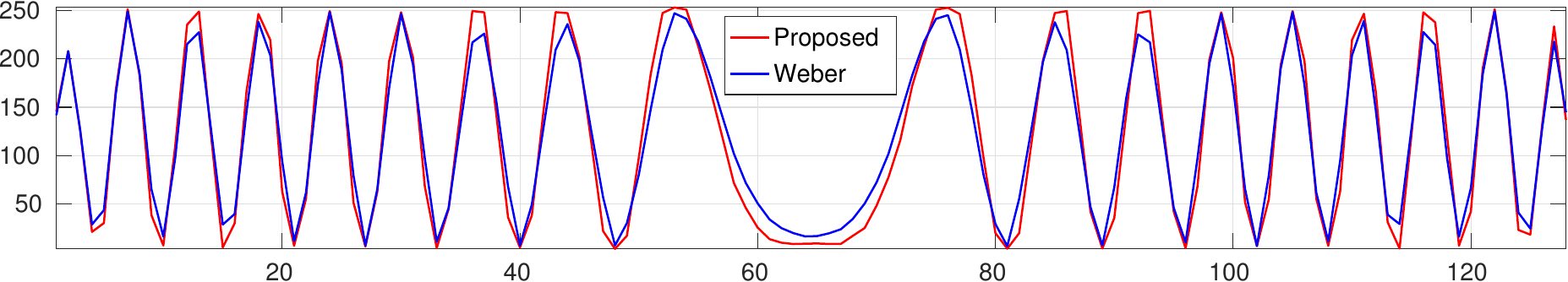}} 
	\\
	\subfloat[The intensities on the input image along the line.]{\includegraphics[width=0.98\linewidth]{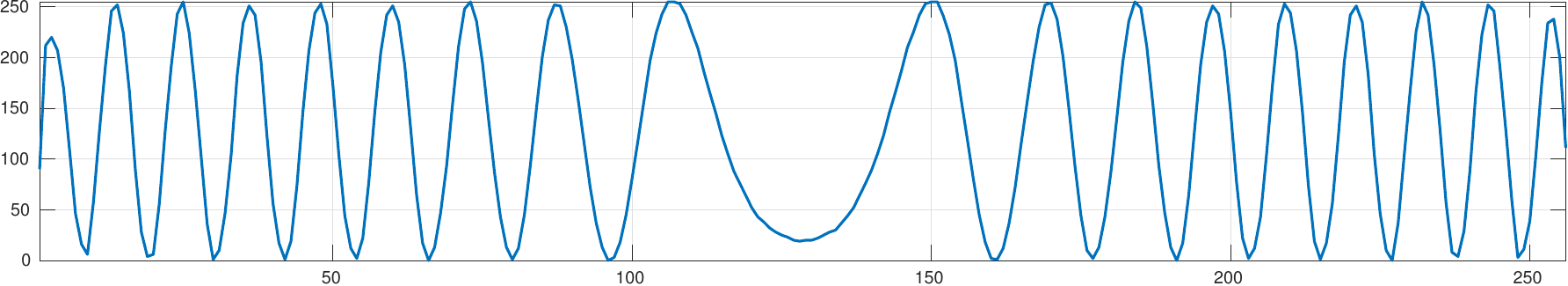}}
	\caption{Downscaling of a radial chirp for a factor $d = 2$. The intensities of the highlighted pixels in both the downscaled images are plotted in Fig. \ref{fig:radialchirp}(d). Notice that our method IDCL can preserve the sharp edges better.} 
	\label{fig:radialchirp}
\end{figure}

\begin{figure}
	\centering
	\subfloat[Input.]{\includegraphics[width=0.56\linewidth]{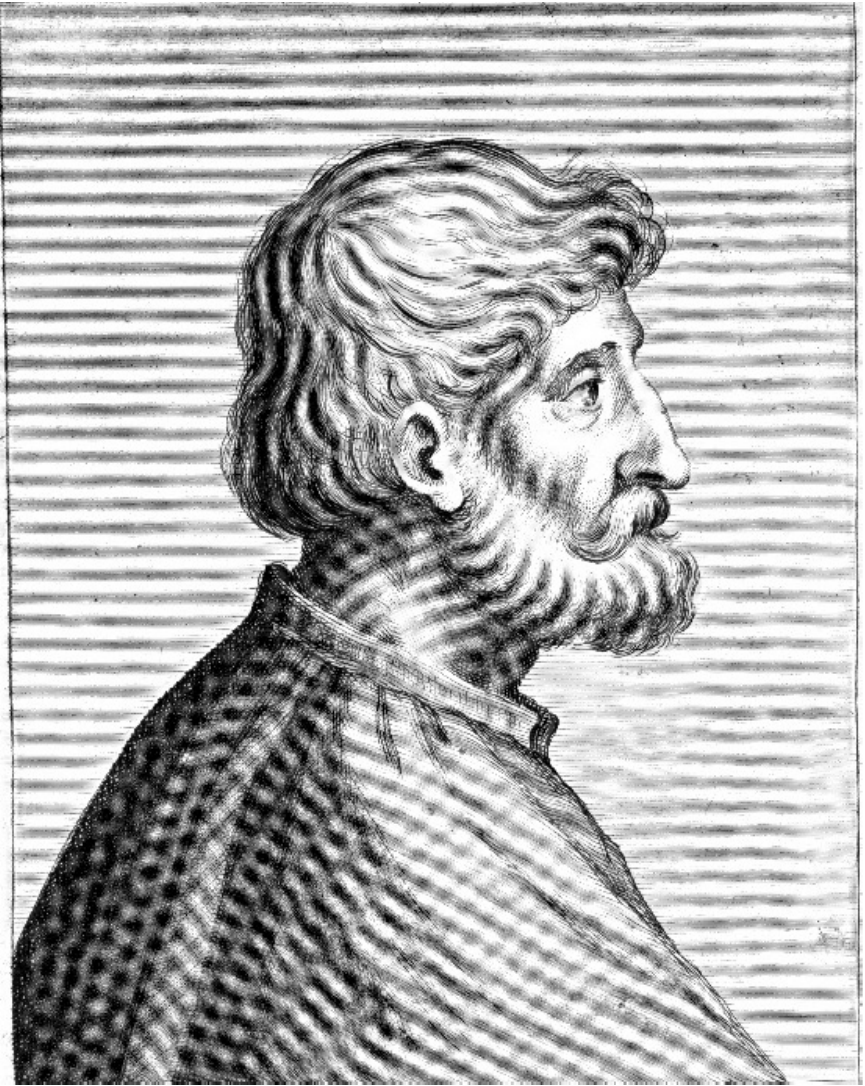}} \\
	\subfloat[k = 2.]{\includegraphics[width=0.28\linewidth]{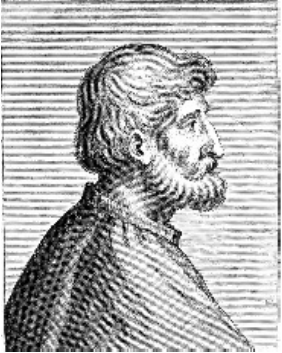}} \vspace{2mm}
	\subfloat[k = 3.]{\includegraphics[width=0.28\linewidth]{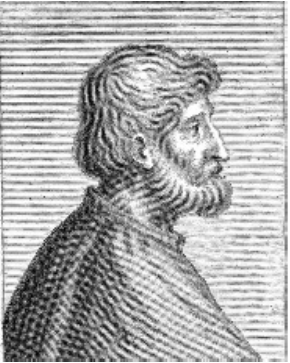}} \vspace{2mm}
	\subfloat[k = 5.]{\includegraphics[width=0.28\linewidth]{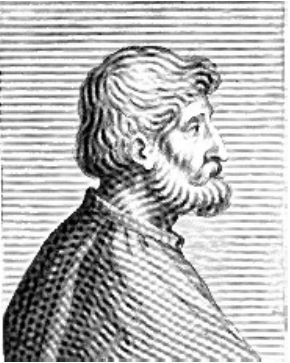}} 
	\caption{Downscaling of a hand-drawn portrait for a factor $d = 3$. The output in (d) obtained using the proposed downscaling technique looks most sharper and clear than than the rest two outputs.} 
	\label{fig:vittore}
\end{figure}
\begin{figure}[h!]
	\centering
	\subfloat[Input.]{\includegraphics[width=0.4\linewidth]{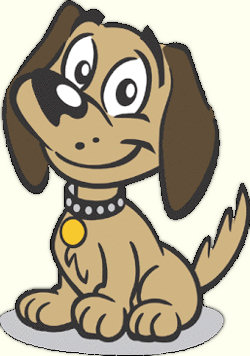}}
	\hspace{0.1mm}
	\subfloat[$d = 2$.]{\includegraphics[width=0.3\linewidth]{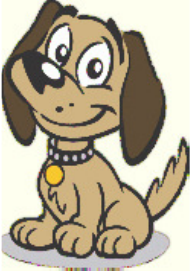}}
	\hspace{0.1mm}
	\subfloat[$d = 4$.]{\includegraphics[width=0.25\linewidth]{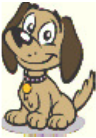}} \\
	\subfloat[Input.]{\includegraphics[width=0.4\linewidth]{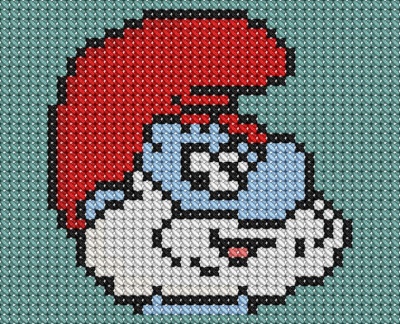}}
	\hspace{0.1mm}
	\subfloat[$d = 2$.]{\includegraphics[width=0.3\linewidth]{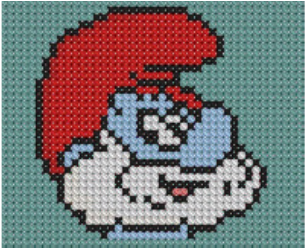}}
	\hspace{0.1mm}
	\subfloat[$d = 4$.]{\includegraphics[width=0.25\linewidth]{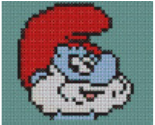}}
	\caption{Downscaling cartoon images using our method. We notice satisfactory performance on both images.}
	\label{fig:cartoon}
\end{figure}

\subsection{Performance on Exhaustive Images}
To validate the practical relevance of our proposal, we performed a large number of experiments on an extensive set of images for many different downscaling factor. We also compare with the existing downscaling methods include subsampling, Lanczos filter \cite{duchon1979lanczos}, bi-cubic, Nehab \cite{nehab2011generalized}, Kopf \cite{kopf2013content}, Oztireli \cite{oztireli2015perceptually}, and DPID \cite{weber2016rapid}. 
In all comparisons with other techniques, the images were either generated using software provided by their own authors or directly downloaded from online repository maintained by the authors. 
We encourage the readers to take zoom-in on the PDF (full-resolution) for best view (and visual comparisons).
The images were obtained from the supplemental/web-page of \cite{kopf2013content}, \cite{weber2016rapid}, and \cite{gastal2017spectral}.
In all the experiments here, we performed uniform downscaling.
Therefore, for a reported downscaling factor $d$, the total number of pixels is reduced by a factor $d^2$.
Here we present only a few representative examples. However, we have covered various different types of practical images as follows:

\begin{table*}[!h]
	\centering
	\caption{Comparion in terms of NIQE metric on different datasets for $d = 2$.}
	\label{table}
	\addtolength{\tabcolsep}{7pt}
	\begin{tabular}{|c|| c |c | c | c|}
		\hline 
		\texttt{Datasets}  & {Lanczos} \cite{duchon1979lanczos}  & {DPID} \cite{weber2016rapid} & \texttt{$L_0$} \cite{liu2018L0} & {IDCL}   \\ \hline 
		\texttt{Set 14}   & 12.23 & 10.68  & 13.53     &10.12   \\ \hline
		\texttt{Urban100} & 7.25   & 6.16  & 8.27	& 5.93 \\ \hline
	\end{tabular}
	\label{tab:niqe}
\end{table*}

\begin{table*}[!h]
	\centering
	\caption{\textcolor{blue}{Comparion in terms of NIQE metric for $d = 4$.}}
	\label{table}
	\addtolength{\tabcolsep}{7pt}
	\begin{tabular}{|c|| c |c | c | c|}
		\hline 
		\texttt{Dataset}  & {Lanczos} \cite{duchon1979lanczos}  & \texttt{$L_0$} \cite{liu2018L0} & VPI \cite{occorsio2022image} & {IDCL}   \\ \hline 
		\texttt{PIXELS300}   & 6.99 & 6.97  & 6.79     &5.57   \\ \hline
	\end{tabular}
	\label{tab:niqe_2}
\end{table*}

\begin{table*}[!h]
	\centering
	\caption{Quantitative evaluation results (PSNR / SSIM) on different datasets for $d = 2$.}
	\addtolength{\tabcolsep}{3pt}
	\begin{tabular}{|c|| c |c | c | c| c|}
		\hline 
		\texttt{Datasets}  & {Lanczos} \cite{duchon1979lanczos}  & {DPID} \cite{weber2016rapid} & {$L_0$} \cite{liu2018L0} & Sun \cite{sun2020learned} & {IDCL}   \\ \hline 
		\texttt{Set 5}   & 35.7 / 98.9 & 41.0 / 99.6  & 26.3 / 93.3   & 34.4 / 94.3  & 43.7/99.7   \\ \hline
		\texttt{Set 14} & 34.4 / 97.7   & 38.4 / 98.9  & 25.3 / 90.8 & 30.2 / 87.1	& 40.3 / 99.2 \\ \hline
	\end{tabular}
	\label{tab:psnr}
\end{table*}

\subsubsection{Synthetic image}
\begin{figure*}
	\centering
	\subfloat[Input.]{\includegraphics[width=0.56\linewidth]{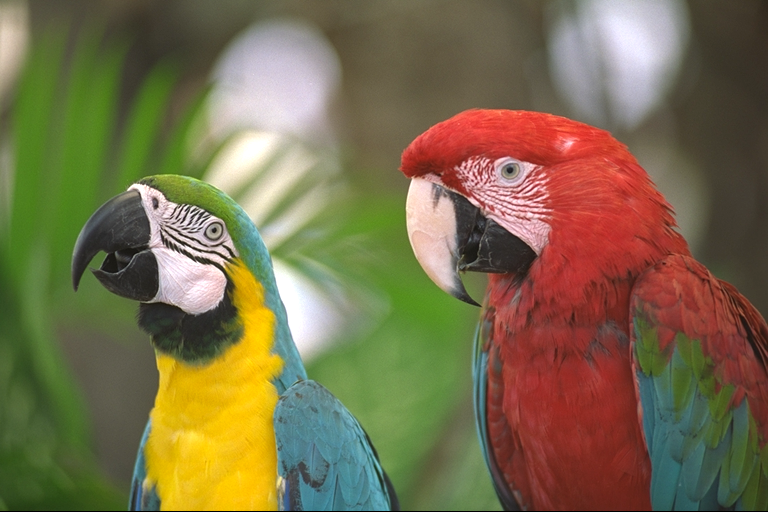}} \vspace{2mm} \subfloat[Close-ups of the red boxes.]{\includegraphics[width=0.4\linewidth]{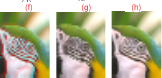}}  \\
	\subfloat[Subsampling.]{\includegraphics[width=0.3\linewidth]{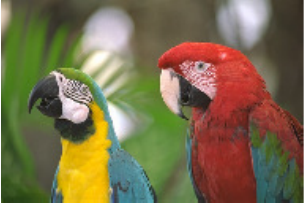}} \vspace{2mm}
	\subfloat[Lanczos.]{\includegraphics[width=0.3\linewidth]{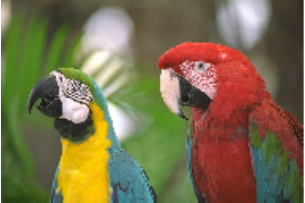}} \vspace{2mm}
	\subfloat[Bicubic.]{\includegraphics[width=0.3\linewidth]{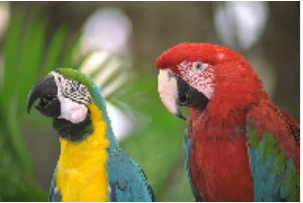}} \\
	\subfloat[Oztireli.]{\includegraphics[width=0.3\linewidth]{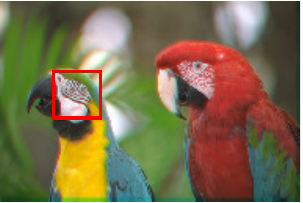}} \vspace{2mm} 
	\subfloat[DPID.]{\includegraphics[width=0.3\linewidth]{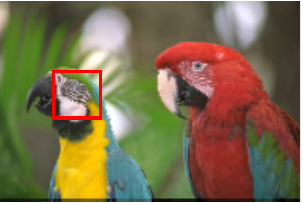}} \vspace{2mm}
	\subfloat[IDCL.]{\includegraphics[width=0.3\linewidth]{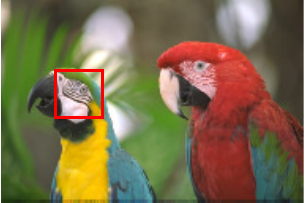}} 
	\caption{Comparison of various image-downscaling techniques. The input image of ($512 \times 768$) pixels is downscaled to ($128 \times 192$) pixels with: (c) subsampling, (d) Oztireli \cite{oztireli2015perceptually}, (e) bicubic,  (f) DPID \cite{weber2016rapid}, (g) Lanczos, and (h) our method IDCL. From the close-ups in (b), it is very evident that the proposed technique outperforms the existing methods. Best viewed at zoomed-in on a computer screen.}  
	\label{fig:birds}
\end{figure*}

An example of downscaling on a synthetic image is shown in Figure \ref{fig:radialchirp}. Here we compare our method with \cite{weber2016rapid}, which was found to be the best (and recent) method as shown in Figure \ref{fig:birds}. 
The blue plot in Figure \ref{fig:radialchirp}(d) is the intensity values of the highlighted pixels in Figure \ref{fig:radialchirp}(b), obtained using \cite{weber2016rapid}. Similarly, the red plot in Figure \ref{fig:radialchirp}(d) is the intensity values of the highlighted pixels in Figure \ref{fig:radialchirp}(c), obtained using our method. We see that the downscaled image using our method IDCL exhibits more details than \cite{weber2016rapid}. In fact, it is desired for an efficient downscaling method to successfully retain the visually important high-frequency details.

\begin{figure*}
	\centering
	\subfloat[Input (512 x 756).]{\includegraphics[width=0.4\linewidth]{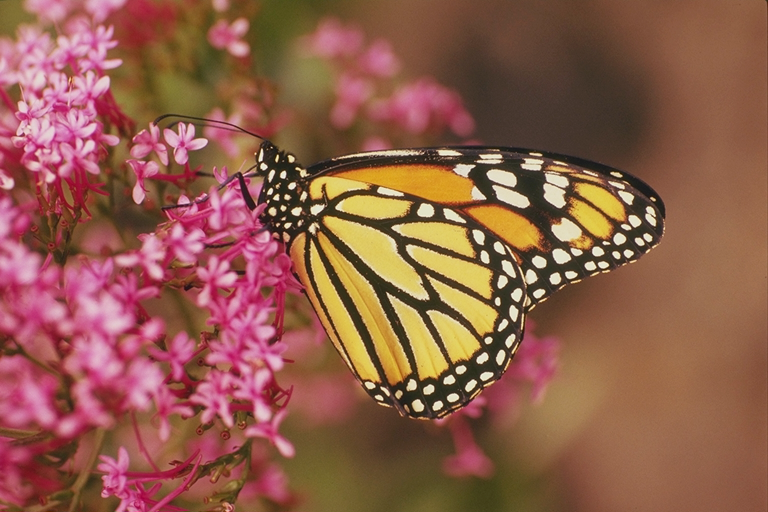}} \hspace{0.1mm}
	\subfloat[$d = 2.3$, NIQE = 4.35.]{\includegraphics[width=0.22\linewidth]{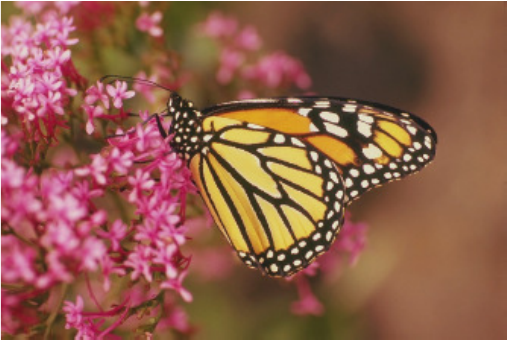}}
	\hspace{0.1mm}
	\subfloat[$d = 3.7$, NIQE = 4.69.]{\includegraphics[width=0.18\linewidth]{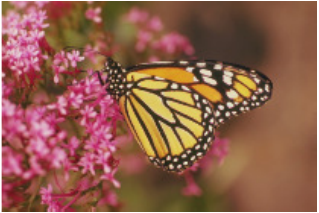}}
	\caption{Downscaling using our method for fractional ratios. The NIQE quality metric is reported. Notice that both downscaled images look visually pleasant.}
	\label{fig:fractional}
\end{figure*}

\begin{figure*}[htp!]
	\centering
	\subfloat[Input.]{\includegraphics[width=0.52\linewidth]{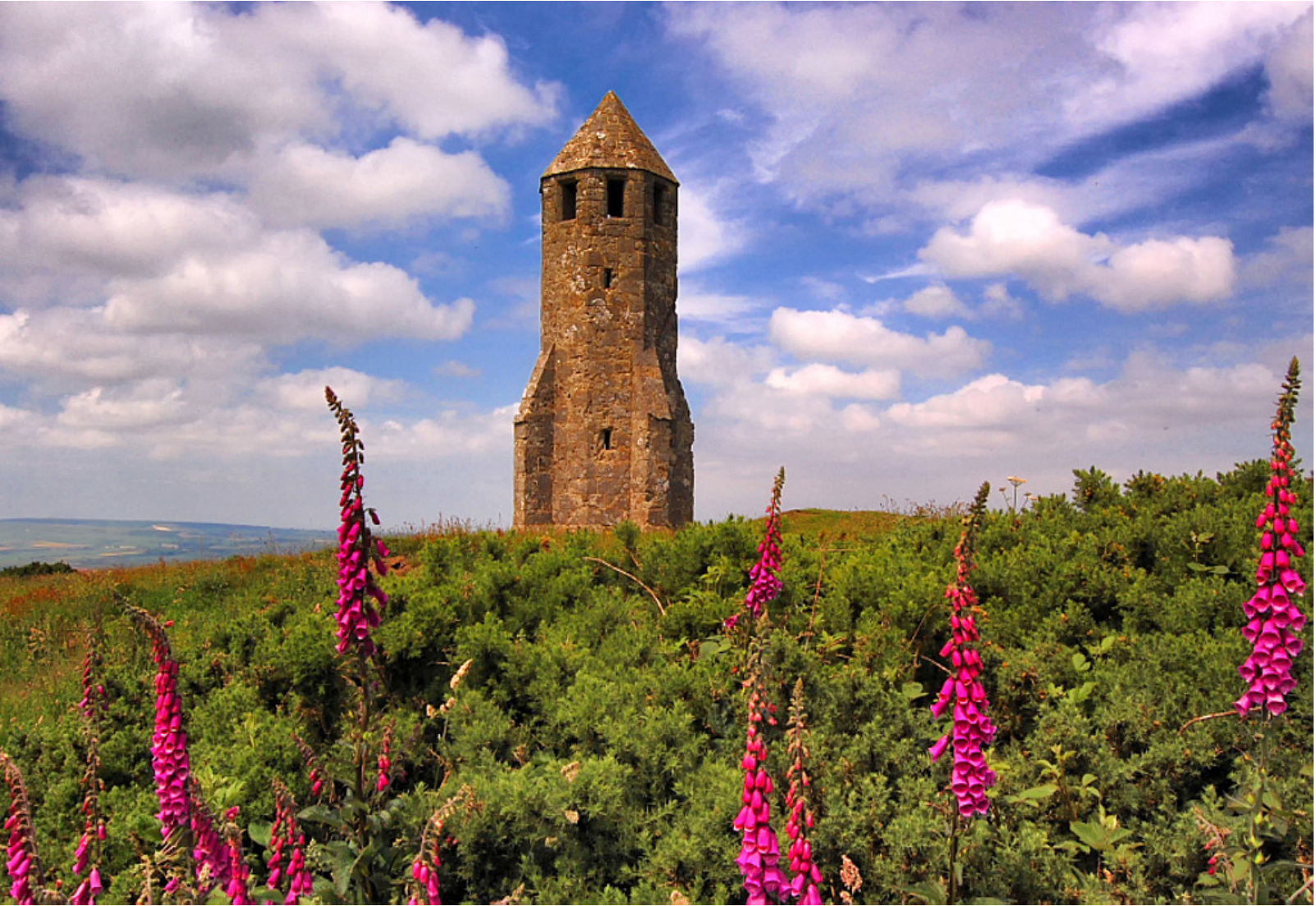}} \vspace{2mm}  \\
	\subfloat[{Hou \cite{hou2017deep}}, \ 5.59.]{\includegraphics[width=0.31\linewidth]{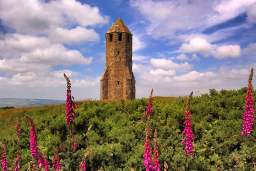}} \vspace{2mm}
	\subfloat[Bicubic, \ 7.07.]{\includegraphics[width=0.3\linewidth]{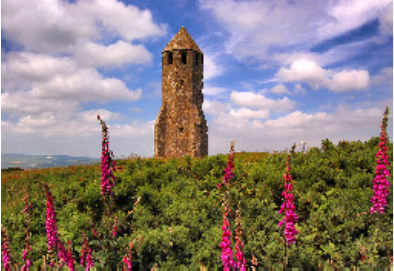}} \vspace{2mm} 
	\subfloat[Lanczos, \ 5.19.]{\includegraphics[width=0.3\linewidth]{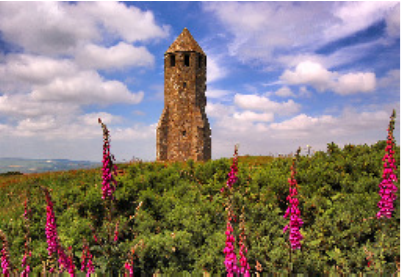}} \\
	\subfloat[Oztireli, \ 6.66.]{\includegraphics[width=0.3\linewidth]{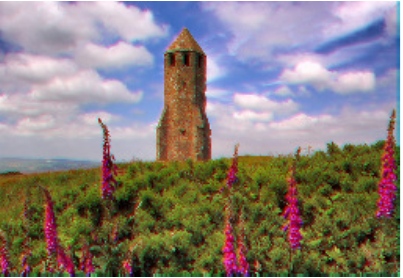}} \vspace{2mm}
	\subfloat[DPID, \ 4.52.]{\includegraphics[width=0.3\linewidth]{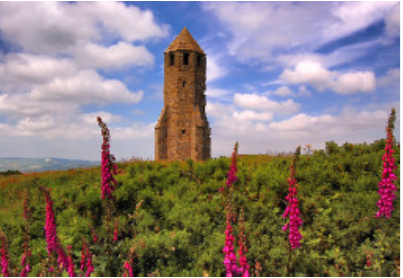}} \vspace{2mm}
	\subfloat[IDCL, \textbf{3.89}.]{\includegraphics[width=0.3\linewidth]{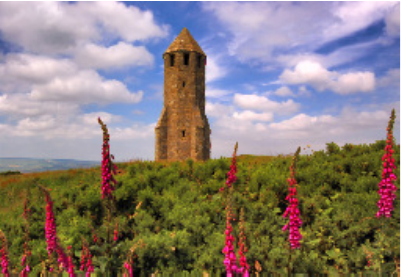}} 
	\caption{Comparison of various image-downscaling techniques. The input image of ($704 \times 1024$) pixels is downscaled to ($176 \times 256$) pixels with: (b) subsampling, (c) bicubic, (d) Lanczos, (e) \cite{oztireli2015perceptually}, (f) \cite{weber2016rapid}, and (g) the proposed method. A no-reference quality metric NIQE \cite{mittal2013making} is mentioned in the captions. A smaller NIQE value represents better image quality. As expected, the input image has the least NIQE value. Among the downscaled images in (b)-(g), the image obtained using the proposed method IDCL has the smallest NIQE value. Best viewed at zoomed-in on a computer screen.}
	\label{fig:tower}
\end{figure*}

\begin{figure*}[htp!]
	\centering
	\subfloat[Input.]{\includegraphics[width=0.52\linewidth]{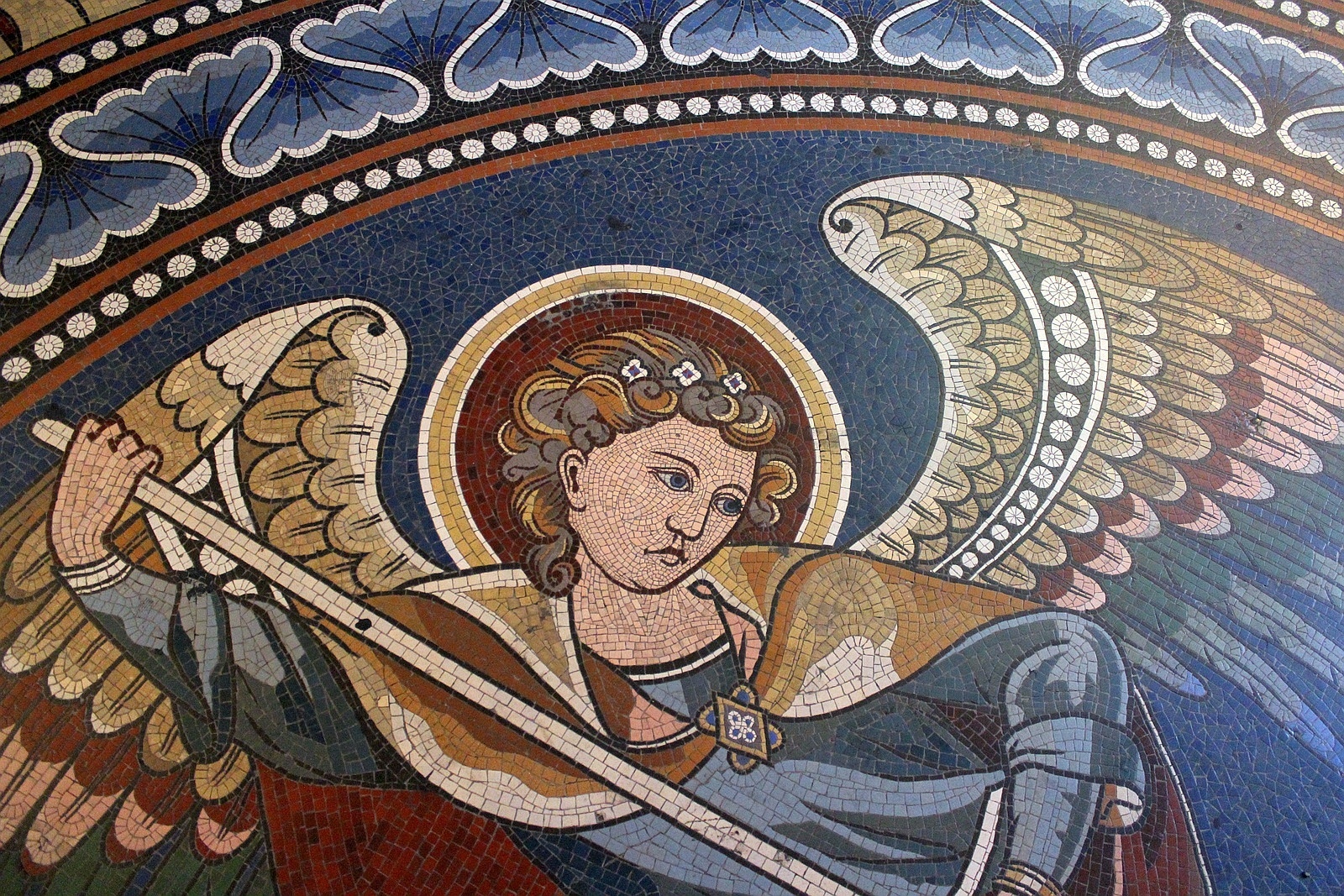}} \vspace{2mm}  \\
	\subfloat[Subsampling.]{\includegraphics[width=0.31\linewidth]{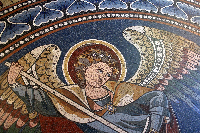}} \vspace{2mm}
	\subfloat[Neheb.]{\includegraphics[width=0.3\linewidth]{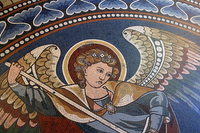}} \vspace{2mm} 
	\subfloat[Kopf.]{\includegraphics[width=0.3\linewidth]{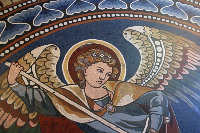}} \\
	\subfloat[Oztireli.]{\includegraphics[width=0.3\linewidth]{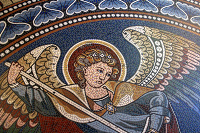}} \vspace{2mm}
	\subfloat[DPID.]{\includegraphics[width=0.3\linewidth]{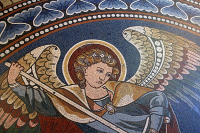}} \vspace{2mm}
	\subfloat[IDCL.]{\includegraphics[width=0.3\linewidth]{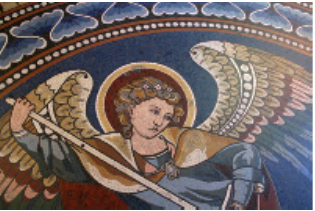}} 
	\caption{\textcolor{blue}{Comparison of various image downscaling techniques for $d = 8$. Best viewed at zoomed-in on a computer screen.}}
	\label{fig:mkichel}
\end{figure*}




\subsubsection{Painted image}
We show the results of downscaling a hand-drawn portrait of the Italian painter Vittore Carpaccio in Figure \ref{fig:vittore}.
The smaller images in Figure \ref{fig:vittore}(b)-(d) contain ($231 \times 184$) pixels; whereas the input image has ($693 \times 552$) pixels. The face in each of the downscaled images is quite instructive. This indicates that there is less blurring in the output of our technique as compared to the rest two methods. In other words, the proposed approach can achieve competitive performance with the existing methods.

{ \subsubsection{Cartoon image}
	In Figure \ref{fig:cartoon}, we show the results on a couple of cartoon images. Two different values of downscaling factor  is considered in this experiment. TWe notice satisfactory performance by our method for both images and scaling factors. }

\subsubsection{Natural image}
We now compare the proposed method with some state-of-the-art methods on natural images.
One such comparison was already reported in Figure \ref{fig:birds}. 
It is evident from the close-ups in Figure \ref{fig:birds}(b) that the proposed method outperforms the other methods in retaining the fine details which were present in the input image. 

In Figure \ref{fig:tower}, we show another comparison on a color image of an out-door natural scene. The downscaling factor is taken as $d = 4$. We notice that the output of our method looks better than the existing approach. However, similar to the previous example, we also use a no-reference quality measure NIQE \cite{mittal2013making} to evaluate the performance of the downscaling methods. 
\textcolor{blue}{Further, we show another visual result for downscaling factor $d = 8$ in Figure \ref{fig:mkichel}.}
Among the output of all downscaling approaches in the list, the one obtained using the proposed method has the least NIQE score; therefore the best downscaling is achieved. We notice in all three figures that our method provides a faithful representation of the input image after downscaling. 
{
	We next performed exhaustive experiments on two larger datasets \cite{sun2020learned} of natural images. The average NIQE values on both datasets are reported in Table \ref{tab:niqe}. \textcolor{blue}{Further, we experimented on another large dataset PIXELS300 \cite{occorsio2022image} in Table \ref{tab:niqe_2} For all three datasets, our method produces better results.}
	We further perform another set of experiments and the results are presented in Table \ref{tab:psnr}. In this experiments, we adapt the experimental protocol as stated in \cite{sun2020learned}. We first upsample each images in a dataset using Matlab ``imresize" function {\color{blue} which uses bicubic interpolation}  and then downscale them using the competing algorithms. Finally, we compuate PSNR and SSIM values between the downscaled and original images \cite{occorsio2022image, occorsio2022lagrange}. Its noteworthy to mention that our proposed method outperforms all others by significant margins.}



\subsubsection{Image with High-Frequency Contents}
\begin{figure}
	\centering
	\subfloat[Input ($400 \times 500$).]{\includegraphics[width=0.52\linewidth]{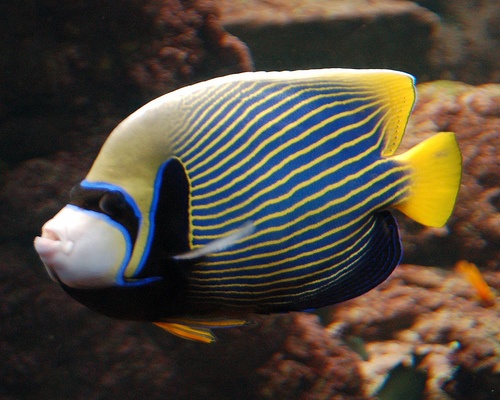}} \vspace{2mm}  \\
	\subfloat[Lanczos.]{\includegraphics[width=0.31\linewidth]{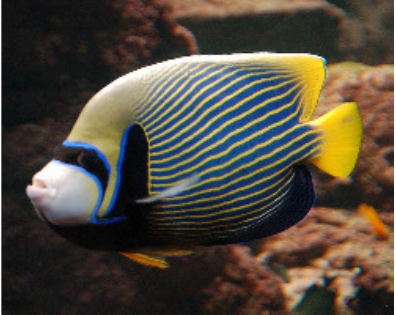}} \vspace{2mm}
	\subfloat[Subsampling.]{\includegraphics[width=0.3\linewidth]{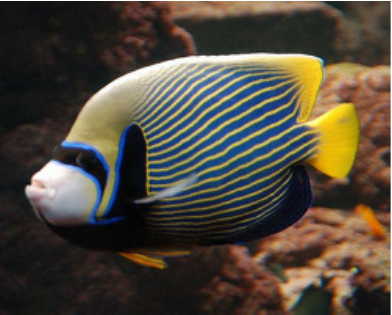}} \vspace{2mm} 
	\subfloat[Kopf.]{\includegraphics[width=0.3\linewidth]{./figures/fish3_subsamp-eps-converted-to.pdf}} \\
	\subfloat[Oztireli.]{\includegraphics[width=0.3\linewidth]{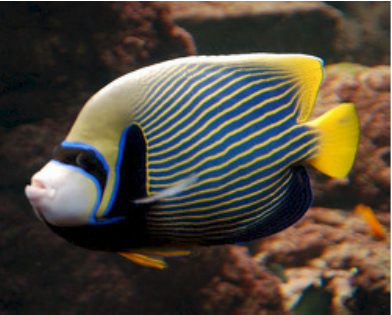}} \vspace{2mm}
	\subfloat[DPID.]{\includegraphics[width=0.3\linewidth]{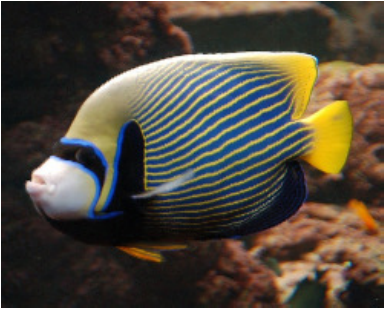}} \vspace{2mm}
	\subfloat[IDCL.]{\includegraphics[width=0.3\linewidth]{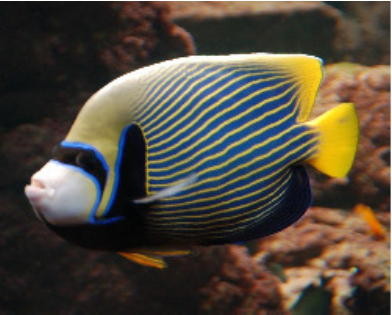} } 
    \caption{\textcolor{blue}{A fish with wiggly stripes downscaled by a factor of $2$.}}
	\label{fig:fish3}
\end{figure}

\begin{figure}
	\centering
	\subfloat[Input ($400 \times 300$).]{\includegraphics[width=0.36\linewidth]{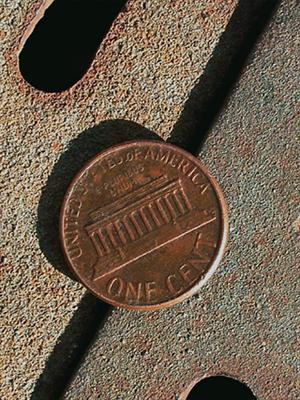}} \vspace{2mm}  
	\subfloat[Oztireli.]{\includegraphics[width=0.18\linewidth]{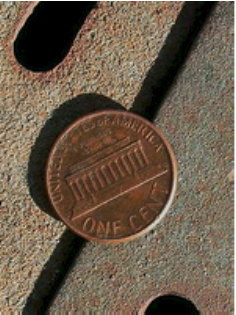}} \vspace{2mm}
	\subfloat[DPID.]{\includegraphics[width=0.18\linewidth]{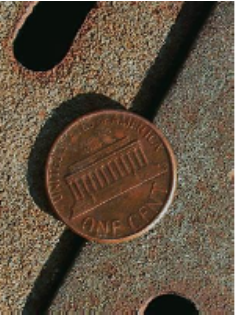}} \vspace{2mm}
	\subfloat[IDCL.]{\includegraphics[width=0.18\linewidth]{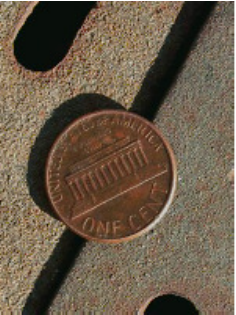} } 
    \caption{\textcolor{blue}{Downscaling (d = 2) of a \textit{coin} image which contains structured fine details.}}
	\label{fig:img06}
\end{figure}

\textcolor{blue}{A typical challenge for most of the downscaling algorithms is to preserve the high-frequency informations present in the input image. To particularly study the performance of our method for this task, we have experimented with the selected images in Figures \ref{fig:fish3} and \ref{fig:img06}.  A fish with wiggly stripes in  Figure \ref{fig:fish3} is downscaled by a factor of $2$. In Figure \ref{fig:img06}, we downscaled a coin image of size of ($400 \times 300$) to an image of ($200 \times 150$) pixels using various techniques. Note that our approach can faithfully retain  various forms of high-frequency fine details in both cases.}

\subsection{Performance for Non-integer Downscaling of Images}
In Figure \ref{fig:fractional}, we show results for fractional factors $d = 2.3$ and $d = 3.7$. Downscaling at fractional ratio is ofter more relevance in practical real-world applications. Notice that our proposed method gives satisfactory downscaling performance for both the fractional factors. We note that unlike some of the existing existing methods, the proposed downscaling approach works for any realistic setting of downscaling.

\subsection{Video Downscaling}
We have performed an experiments on video processing. Both input and downscaled videos are publicly available at \textbf{https://github.com/Sanjay-Ghosh/Image-Downscaling}. Please follow the instruction in README.md for displaying the videos in Matlab. Notice that there is no jittering effects present in the output video.

\section{Discussion}
We propose a new downscaling method IDCL\footnote{Matlab codes are available at \href{https://github.com/Sanjay-Ghosh/Image-Downscaling}{https://github.com/Sanjay-Ghosh/Image-Downscaling}.} by exploring the existence of pair-wise co-occurrence in pixel intensities of natural images.  In fact, this is the key idea used in the co-occurrence learning in Algorithm 1. The parameter $d$ is basically used for restricting the neighborhood during the co-occurrence learning. The term "non-local" refers the fact that for a particular pair of image intensities $(f(i), f(j))$, we do search within the whole image under the restriction that pixel $i$ lies in the neighborhood of pixel $j$. This neighborhood size is determined by parameter $d$.

The proposed method is an instance of histogram consistency \cite{lugosi1996consistency}. The pairwise co-occurrence tends to maintain a similar pairwise histograms between the input and downscaled images. We consider the task of a theoretical analysis  as a potential future exercise.
We believe that our work has potential in many aspects. This co-occurrence learning could be useful in improving feature learning in neural networks. Unlike many existing approaches, our method works for fractional ratios. We humbly draw your attention that the proposed method outperforms (in terms of PSNR/SSIM) the most recent methods  \cite{liu2018L0, sun2020learned} by a good margin.

\subsection{Limitations}
\begin{figure}
	\centering
	\subfloat[Input.]{\includegraphics[width=0.60\linewidth]{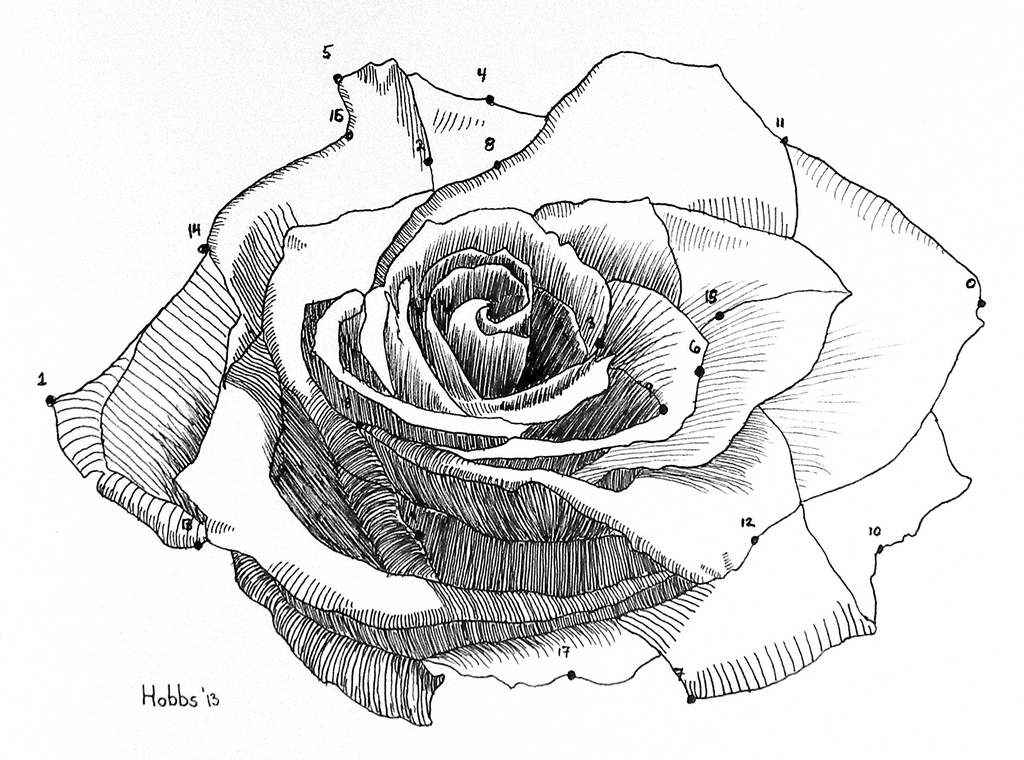} } \hfill 
	\subfloat[IDCL.]{\includegraphics[width=0.35\linewidth]{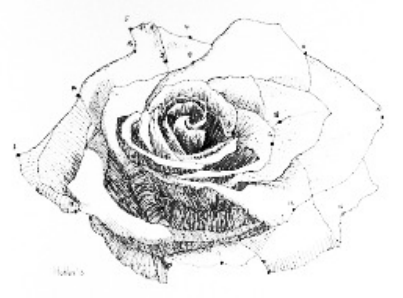} } 
    \caption{\textcolor{blue}{Rose image is downscaled from ($760 \times 1024$) to ($190 \times 256$) pixels using $d = 4$.}}
    \label{fig:rose}
\end{figure}

\textcolor{blue}{In general, the proposed technique produces superior results compared to the existing techniques. It was indicated by the participants in the user study that our method generates perceptually accurate and appealing results.  However, it may occasionally smooth out fine details while downscaling certain synthetic or drawing images. One such example is shown in Figure \ref{fig:rose}. In this experiment, the input image of ($760 \times 1024$) pixels is downscaled to ($190 \times 256$) pixels.
Our algorithm basically relies on the heuristic that all possible pixel intensity pairs are sort of fairly distributed over the non-local neighborhood used for building the co-occurrence matrix $h$. Therefore, the co-occurrence frequency of each pair is assumed to be a reliable feature to be used for the similarity measure in the filtering step. However, this simple assumption does not hold for the input image in this example. Finding a better feature (beyond co-occurrence frequency) for this kind of challenging downscaling task would be an interesting future exercise. }

~\\
\textbf{Software}: \\
{https://github.com/Sanjay-Ghosh/Image-Downscaling}

\section{Conclusion}
\label{sec:conc}
In this paper, we proposed a new method for image downscaling. The core idea is a special instance of kernel filtering, where pixel-pair co-occurrence was used as the similarity measure. The co-occurrence feature was directly learned directly from the input image. In a perspective, the proposed feature learning based approach opens up a new direction for learning based downscaling in future.
We showed that the proposed image downscaling technique can achieve better performance than existing methods
qualitatively as well as quantitatively. 
Despite of its effective downscaling capacity, it has very simple and parallelizable implementation. {\color{blue} The algorithm can be applied to downscale images and videos in real-world applications such as social media applications to display on smaller devices as well. The LR images help quicker broadcast and take less storage.
}




\bibliography{citations_downscaling.bib}

\end{document}